\documentclass[pra,showkeys,showpacs]{revtex4}

\usepackage{graphicx} 

\newcommand{\be}{\begin{equation}}
\newcommand{\ee}{\end{equation}}
\newcommand{\bea}{\begin{eqnarray}}
\newcommand{\eea}{\end{eqnarray}}

\newcommand{\nn}{\mbox{$\nonumber$}}

\newcommand{\zit}{\int_{-\infty}^{t}}

\newcommand{\vA}{\vec{A}}
\newcommand{\vB}{\vec{B}}
\newcommand{\vE}{\vec{E}}
\newcommand{\vdl}{\vec{d \ell}}
\newcommand{\vm}{\vec{\mu}}
\newcommand{\vp}{\vec{p}}
\newcommand{\vv}{\vec{v}}
\newcommand{\vF}{\vec{F}}
\newcommand{\ver}{\vec{r}}

\begin{document}

\title{\bf{ Derivation of the Aharanov Bohm phase shift using classical forces.  }}
\author{\small H. Fearn  and Khai Nguyen \\
\em \small Physics Department\\
\em \small California State University Fullerton\\
\em \small 800 N. State College Blvd., Fullerton CA 92834\\
\small phone (657) 278 2767\\
\small email hfearn@fullerton.edu }

\begin{abstract}
In 1959 Aharonov and Bohm suggested that an electron passing around a long solenoid would pick up a
phase shift dependent on the magnetic field of the solenoid, even though the electrons themselves
pass through a region of space which has a zero magnetic field. It has long been held that this result
is purely quantum and is derived in many well known quantum mechanics text books using the Schrodinger equation
and vector potential. Here the same phase shift is derived from a purely classical force, but relativistic
transformations are taken into account.
The force is in the direction of motion of the electron (or opposite) leading to a phase advance (or lag)
and we obtain precisely the phase shift thought previously to be purely quantum. The only quantum result
used here is the de Broglie wavelength of the particle, in order to get two slit like
interference and the phase shift. We employ a stack of dipoles as the solenoid and note the
same force on the electron in two different frames of reference. We shall consider the solenoid stationary and the
electron moving, and then consider the electron rest frame and consider the solenoid moving in the
opposite direction.
\end{abstract}



\maketitle

\vspace{0.25in}

\subsection*{1. Introduction}

\noindent
The Aharanov Bohm (AB) effect \cite{AB}, is a well known example in quantum text books of a
``surprising'' phase shift for charged particles passing around a long thin solenoid. The charged particles pick up
an unexpected phase shift proportional to the magnetic field of the solenoid even though the particles
experience no magnetic field on their path around the solenoid. Many related experiments have been proposed
where the phase shift is ``clearly'' due to a classical force. For example there are experimentally observed effects with
gravitational fields \cite{grav} and electrostatic fields \cite{electro}. More recently Boyer \cite{boyer} has shown
that the Aharanov Casher effect \cite{AC}, which describes a magnetic dipole moving around a line charge, can be described by a
purely classical force. In this paper, we show that a very similar type of derivation, to that of Boyer above, is
possible for the AB effect,
and we claim that there is no mysterious quantum shift, the phase is due to a classical force just like the
other observed effects. We note that Boyer has suggested that the AB effect does have a velocity lag/gain
effect in an earlier paper, \cite{boyer2}, although his derivation involves a WKB quantum calculation, quite
different from what we have here. A later paper also by boyer \cite{boyer3} has a rather similar calculation although
 not involving a relativistic transformation which we use here. I have been interested in these effects for some time and have previously published a
related article on an optical analogue to the AB-effect with Cook and Milonni, \cite{fearn}. In that paper the
reader may find several other analogues to the AB effect.  We will now continue with the text book account and
then show the classical derivation in S.I. units.

\noindent

\vspace{0.25in}

\subsection*{2. The usual AB effect derivation via the Schrodinger Equation. }

\noindent
For completeness I will give the starting equations and final result for the usual text book approach to the
AB effect and a brief explanation of what the phase shift should look like. Sample text books include ``Advanced
quantum mechanics'' by Sakurai \cite{sak}, and an ``Introduction to quantum mechanics'' by
Griffiths \cite{griff}.\\

 Since the only force in these type of problems is along the direction of
motion of the moving particle or directly opposite, there is no overall shift of the interference pattern,
only a small relative shift of the 2 slit interference under the single slit diffraction pattern envelope.
We have a picture of this in the optical analogue paper \cite{fearn}.
The single slit diffraction envelope does not move, if you were to take away the solenoid the peak in the
single slit diffraction pattern would match the peak in the interference pattern, the interference peak
shifts slightly to one side when the solenoid is placed in position behind the 2 slits, or between to
2 paths of the particle.\\

\noindent
The standard text book approach involves the time-dependent Schrodinger equation for an electron
in a vector potential, using S.I. units,

\be
\frac{1}{2m} \left( \frac{\hbar}{i} \nabla - e \vA \right)^2 \psi = E \psi
\ee

which can be expanded out as,
\be
\nabla^2 \psi - \frac{ie}{\hbar } \left[ \vA \cdot \nabla \psi + (\nabla \cdot \vA )\psi \right]
+ \left( \frac{2mE}{\hbar^2} + \frac{ e^2}{\hbar^2} \vA^2 \right) \psi =0 .
\ee

If we take $ \psi_0 (r) $ as the solution when $ \vA =0 $, then a solution with $ \vA \neq 0$ is,

\be
\psi (r) = \psi_0 (r) \exp \left[ \frac{ie}{\hbar } \int_{\mbox{\small path}} \vA (r') \cdot \vdl' \right]
\ee

Now consider the electron double slit experiment, where we assume the electron only travels along a path where
we may set $ \vB = \nabla \times \vA =0$. The interference pattern is produced by the cross term in $ |\psi(r) |^2$
 which is twice the real part of,
 \be
 \psi_{1} (r) \psi_{2}^\ast (r) = \psi_{01} \psi_{02}^\ast \exp \left[
 \frac{ie}{\hbar } \left( \int_{\mbox{\small path 1}} \vA (r') \cdot \vdl' -
\int_{\mbox{\small path 2}} \vA (r') \cdot \vdl'
 \right) \right]
 \ee

 where $ \psi_{oi}(r) $ is a solution of the Schrodinger equation when $ \vA =0$ and only one slit is present and
 subscripts refer to slit 1 or 2. Using Stokes' theorem and $ \vB = \nabla \times \vA $, we can write the
 this interference term as,

 \be
\psi_{01} \psi_{02}^\ast \exp \left[ \frac{ie}{\hbar }  \int \vB \cdot d S  \right]
\label{wow}
 \ee
where $d S$ is the area bounded by the two paths of the electron. The double slit interference pattern is
therefore modified by the magnetic field even though the electrons paths are confined to regions
where $ \vB =0 $.  We shall refer back to the main result \ref{wow} later.\\

\noindent
In this paper we shall consider the AB effect from two frames of reference.  The solenoid rest frame and
the electron rest frame. Both of these frames yield the same force exerted on the electron. Relativistic
transformations will be used for the electric and magnetic fields although we shall take
$\gamma = ( 1 - v^2/c^2 )^{-1/2} \simeq 1 $ and $ v^2 \ll c^2 $ where $v$ is the electron speed and $c$ is the
velocity of light.

\vspace{0.25in}

\subsection*{3. Classical force from the perspective of the solenoid rest frame. }

\noindent
We will treat the solenoid as a stack of very small dipoles. We orient the stack so that the resulting
magnetic dipole moment of the solenoid $\vm = \mu \hat{i} $ is along the x-axis. The solenoid is at $x=z=0 $
and $y= \pm y_0 $. The sign of $y$ depends on whether we want the electron to pass the solenoid on the left
or the right.
The electrons will move along the z-axis at $ x=y=0$ with velocity $ \vv = v_0 \hat{k} $. See Figure 1.

\vspace{0.25in}

\noindent
\begin{figure}
  \includegraphics[width=3.0in]{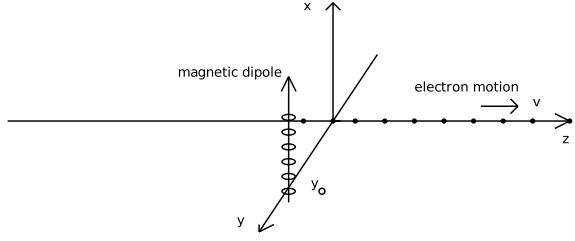}\\
 \caption{Arrangement of the solenoid and path of the  electron.}\label{Fig. 1.}
\end{figure}


\noindent
We know that the phase shift is very small, so we will assume that the electron has a negligible acceleration
in the forward direction.
There will be an acceleration, but for a short time and of very small magnitude. We neglect
gravitational acceleration here also as being too small.  This calculation looks very similar to one
performed by Boyer nearly ten years ago \cite{boyer3}. The AB phase shift has also been experimentally confirmed by
Caprez et al \cite{batelaan,boyer4}.

According to classical electromagnetism,
the electric and magnetic fields of an electron with charge -e moving with constant velocity will be \cite{griff2},
\bea
\vE &=& \frac{ -e}{ 4 \pi \epsilon_0 \gamma^2 } \frac{1}{( 1 - \frac{v^2}{c^2} \sin^2 \theta )^{3/2}}
\frac{ \hat{r}}{r^2} \nn \\
\vB &=& \frac{1}{c^2} ( \vv \times \vE )
\eea

where the $ \theta$ is measured from the forward direction, in this case from the z-axis. We are mainly interested
in the magnetic field from the electron since this will interact with the magnetic dipole moment of the solenoid.
The energy of interaction will be,
\be
\varepsilon = - \vm \cdot \vB ,
\ee
and this leads to a force on the electron of,
\be
\vF = \nabla ( \vm \cdot \vB ) .
\ee
We take $\ver = y \hat{j} + z \hat{k} $ and $r^2 = y^2 + z^2 $. Most of the electromagnetic field comes from
when $\theta = \pi/2$ and so the velocity field from the electron is predominantly perpendicular to the
direction of motion, for a slow electron.  We assume the electromagnetic radiation takes a direct path from
the electron to the solenoid hence set $x=0$.
We can simplify the electric field and hence the magnetic field to be,
\bea
\vE &=& \frac{ -e \gamma}{ 4 \pi \epsilon_0 }\frac{ \hat{r}}{r^2} \nn \\
\vB &=& \pm \frac{e v_0 y}{ 4 \pi \epsilon_0 c^2} \frac{ \hat{i}}{ ( y^2 + z^2 )^{3/2} }
\eea
where we have used $\vv = v_0 \hat{k} $ for the electron.  The $\pm$ tells us whether the electron passes the
solenoid on the left or the right since we can change the position of the solenoid $ y = \pm y_0 $ .
 Since the B-field changes sign then so will the direction of the force. The classical force on the electron,
 due to the magnetic dipole and magnetic field interaction is,
\bea
\vF &=& \pm \left( \frac{\partial}{\partial x} \hat{i} +\frac{\partial}{\partial y} \hat{j} +
\frac{\partial}{\partial z} \hat{k} \right)\left(
\frac{e \mu v_0 y}{ 4 \pi \epsilon_0 c^2} \frac{1}{ ( y^2 + z^2 )^{3/2} } \right) \nn \\
 &=& \pm \frac{e \mu v_0 y}{ 4 \pi \epsilon_0 c^2} \left[
 \frac{-3 z y \hat{k}}{( y^2 + z^2)^{5/2}}
+ \frac{ \hat{j}}{ (y^2 + z^2 )^{3/2} } - \frac{-3 y^2 \hat{j}}{( y^2 + z^2)^{5/2}}  \right]
\label{force}
\eea

From here we may calculate the small change in velocity that the electron would experience from,
\be
\triangle \vv = \zit \frac{\vF}{m} dt'
\ee
we take $dt' \approx dz'/v_0 $  and then find the small forward or backward shift in position
$\triangle z$ due to this interaction,

\be
\triangle z = \zit \triangle \vv \;\; dt'  .
\ee
The integrations can be performed by hand using a substitution of the form $ z = y_0 \tan u$ and
$dz = y_0 \sec^2 u \; du $. The use of Pythagoras will be helpful in evaluating the limits of the integrals.
Form a triangle with perpendicular axes $y_0$, $z$ and tilted axis at angle $\theta$.


Finally using the de Broglie wavelength $ \lambda = h/ ( m v_0 ) $ we obtain the AB phase shift $\phi$ as,
\bea
\phi &=& \frac{ 2 \pi}{\lambda} 2 \triangle z  \nn \\
&=& \frac{ e \mu }{ \hbar \pi \epsilon_0 c^2 y_0 }
\eea
we use $ 2 \triangle z$ since the shift is a lag along one path and gain in the other.
So, how does this compare with the original AB effect phase shift in Eq. (\ref{wow}) ?
\be
\phi_{\mbox{\tiny AB}} = \frac{e}{\hbar} B A
\ee
where the area $A$ is in the yz-plane. Comparing the AB effect phase $\phi_{\mbox{\tiny AB}}$ with the result
here, for perfect agreement, we require that the solenoid magnetic field times the area of interaction is,
\be
 B A = \frac{\mu_0 \mu}{ \pi y_0 }
\ee

where $ \mu_0 \epsilon_0 = 1/c^2 $ has been used. The magnetic field from a static solenoid (or stack of dipoles
in our case) oriented in the x-direction would be, see for example Griffiths' text on E\&M \cite{griff2},
\bea
\vB_{\mbox{\tiny dip}} &=& \frac{\mu_0 }{4 \pi r^3} \left[
                  3( \vm \cdot \hat{r} ) \hat{r} - \vm \right]  \nn \\
 &=& \frac{ \mu \mu_o }{ 2 \pi r^3} \hat{i}
\eea

where the solenoid is oriented along the x-axis  and located at $z=0$ and $ y = \pm y_0$. If we set $r=y_0$,
which is the distance of closest approach of the electron to the solenoid,
and take $2y_0^2$ as an approximate interaction area $A$ then the results match perfectly.
The AB effect shift was first measured experimentally by Chambers, \cite{chambers}.
 The $r$ would in any event lie in the yz-plane since this is the plane of motion
of the electron.

\subsection*{4. Classical force from the perspective of the electron rest frame.}

In this section we consider the electron as stationary and therefore having a static electric field,
and the solenoid (or stack of dipoles) as moving with velocity $\vv = - v_0 \hat{k} $. A magnetic dipole
will appear to have the properties of an electric dipole when it is moving \cite{boyer,jackson}.
We may set the electric dipole moment of the solenoid to be,
\be
\vp = \frac{\vv}{c^2} \times \vm = \frac{ v_0 \mu}{ c^2} \hat{j}
\ee
\noindent
where we have used $ -\hat{k} \times \hat{i} = \hat{j}$ .
The electron has the electric field,
\be
\vE = \frac{ -e \ver }{ 4 \pi \epsilon_0 r^3 }
\ee
and so we have the standard interaction energy $ \varepsilon = - \vp \cdot \vE $. This leads to a force
on the electron of,
\bea
\vF &=& ( \vp \cdot \nabla )\vE = \left(\frac{ v_0 \mu}{ c^2} \hat{j} \right) \cdot
\left( \frac{\partial}{\partial x} \hat{i} +\frac{\partial}{\partial y} \hat{j} +
\frac{\partial}{\partial z} \hat{k} \right) \vE   \nn \\
&=& \frac{ v_0 \mu}{c^2} \;\; \frac{\partial \vE}{\partial y }
\eea
since all motion is in the yz-plane, it is fine to set $ \ver = y \hat{j} + z \hat {k} $ and $r^2 = y^2 + z^2$ here.
We find exactly the same force as in the previous section, see Eq. (\ref{force}). The rest of the calculation
follows exactly as before and of course we obtain the same phase shift $\phi$.

\vspace{0.25in}

\subsection*{5. Conclusions}

\noindent
We have shown that it is possible to derive a realistic Aharonov Bohm phase shift from purely classical forces.
These forces result from a magnetic interaction in one frame and an electric interaction in another but the
resulting force on the electron is the same in both frames of reference. We have made use of the electron velocity
field, which would be seen as a magnetic field by the solenoid as the electron moves by.
This magnetic field is the cause of the interaction force on the magnetic
dipole of the solenoid. In the frame where the electron is at rest, we have the static electric field of the electron
interacting with the electric dipole moment of the moving solenoid. Treated in a fully relativistic fashion
the dipole fields would be part of a second rank tensor, \cite{corben}. We have
assumed a slow motion for the electron so all $\gamma$ factors have been set equal to one.

In the final step of the calculation we had to allow for an area in order to calculate the phase shift. We took the
smallest area the electron paths would surround using the $y_0$ as the distance of closest approach of the
electron to the solenoid. We show a figure below which shows the area we used of magnitude $ 2 y_0^2$, which is
the area enclosed in the box of side length $ \sqrt{2} y_0$. The blue dots show the electron path, and the
ellipses show the stack of dipoles we used as the solenoid, which is now centered.\\

The result here compliments
the work done by Boyer, \cite{boyer3}. The aim was to bring attention to a 30 year old debate on this subject
 and the plight of experimentalists who may still be trying to show the differences between a purely quantum phase
 shift and a semiclassical one. We believe that up to this time there is no absolute proof that the AB phase shifts
seen so far are purely quantum in nature as the experiments done to date are not conclusive.
This may change in the near future however.\\

\noindent
\begin{figure}
  \includegraphics[width=3.0in]{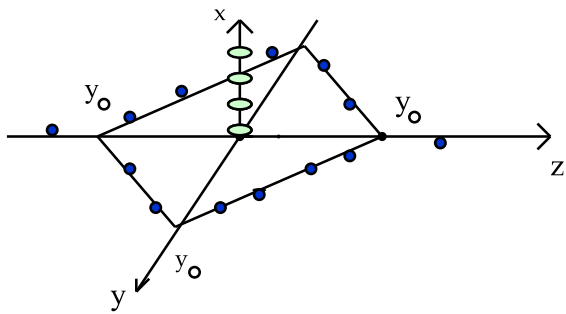}\\
 \caption{Minimum Area enclosed by the electron path.}\label{Fig. 2.}
\end{figure}

\vspace{0.25in}

\subsection*{Acknowledgements}

The inspiration for this work came from teaching my graduate
electromagnetism and quantum classes. We went over the AB-effect
and the AC-effect and worked through the paper by Boyer
\cite{boyer}. This gave me the inspiration to rework the AB-effect
in the same fashion. I would also like to thank Prof. Milonni  and Prof Boyer for emailed comments
 and suggestions.



\vspace{0.5 in}

\begin{thebibliography}{99}

\bibitem{AB} Y. Aharanov and D. Bohm, Phys. ev. {\bf 115}, 485 (1959).

\bibitem{grav} R. Collella, A. W. Overhauser and S. A. Werner, Phys. Rev. Lett. {\bf 34} 1472 (1974).

\bibitem{electro} S. Matteucci and G. Pozzi, Phys Rev. Lett. {\bf 54}, 2469 (1985).

\bibitem{boyer} T. H. Boyer , Phys. Rev. A. {\bf 36}, (10) 5083 (1987).

\bibitem{AC} Y. Aharanov and A. Casher, Phys. Rev. Lett. {\bf 53}, 319 (1984).

\bibitem{boyer2} T. H. Boyer, Phys. Rev. D {\bf 8}, (6) 1679 (1973).

\bibitem{boyer3} T. H. Boyer, Found. of Phys. {\bf 32}, (1) p41 (2002). See also the calculation with a
point charge and a magnetic dipole in, Found. of Phys. {\bf 32} (1) p1 (2002).

\bibitem{fearn} R. J. Cook, H. Fearn and P. W. Milonni , Am. J. Phys. {\bf 63} (8) 705 (1995).

\bibitem{sak} J. J. Sakuari, ``Advanced Quantum Mechanics", (Addison-Wesley,  Reading, MA, 1976) pp. 15-17.

\bibitem{griff}D. J. Griffiths, ``Introduction to quantum mechanics", 2nd Ed. (Prentice Hall, NJ 2005) pp384-391.

\bibitem{batelaan} A. Caprez, B. Barwick and H. Batelaan, Phys Rev. Lett. {\bf 99}, 210401 (2007).

\bibitem{boyer4} T. H. Boyer, Found. of Phys. {\bf 38}, p498 (2008).

\bibitem{griff2}D. J. Griffiths, ``Introduction to Electrodynamics", 3rd Ed. (Prentice Hall, NJ 1999) pp439-440.

\bibitem{chambers} R. G. Chambers, Phys. Rev. Lett. {\bf 5}, p3 (1960). See ref \cite{boyer3} for more experimental
references.

\bibitem{jackson} J. D. Jackson, ``Classical Eectrodynamics'' 3rd Ed. (John `Wiley \& Sons, Berkeley 1999). See
page 292, problems 6.21, 6.22, and page 577 problems 11.28 and 11.29. (In the 1st Ed. p389 problem 11.9.)

\bibitem{corben} H. C. Corben, ``Classical and Quantum Theories of Spinning Particles" ,(Holden-Day, San
Francisco, 1968), p69 Eq., (6.18).



\end{thebibliography}
\end{document}